\documentclass[twocolumn,preprintnumbers,prl,aps,superscriptaddress]{revtex4}

\usepackage{latexsym}          % essential
\usepackage[dvips]{graphicx}   % essential
\usepackage{amsmath}           % essential

\usepackage{amsfonts}          % AMS fonts
\usepackage{amssymb}           % AMS symbols

\usepackage[english]{babel}

\newtheorem{theorem}{Theorem}
\newtheorem{lemma}{Lemma}

\newcommand{\be}{\begin{equation}}
\newcommand{\ee}{\end{equation}}
\newcommand{\ef}[1]{\, #1}

\newcommand{\mysf}[1]{$\mathsf{#1}$}
\newcommand{\sht}[1]{{#1}-SHT}

\newcommand{\Pf}{\mathrm{Pf}\,}
\newcommand{\GF}{\mathrm{GF}}
  \newcommand{\BF}{\mathrm{GF}}

\def\psibar{{\bar{\psi}}}

\begin{document}
% ======================================================= %

%%%%%%%%%%%%%%%%%%%%%%%%%%%% TITLE
\title{A randomized polynomial-time algorithm\\
for the Spanning Hypertree Problem on 3-uniform hypergraphs}

%%%%%%%%%%%%%%%%%%%%%%%%%%%% AUTHORS
\author{Sergio Caracciolo}
\affiliation{Dip.~Fisica,
  Universit{\`a} degli Studi di Milano, and INFN, via G.~Celoria 16,
  20133 Milano, Italy}
\author{Gregor Masbaum}
\affiliation{Institut de Math{\'e}matiques de Jussieu (UMR 7586 CNRS),
  Universit{\'e} Paris Diderot, Case 7012 - Site Chevaleret, 75205 Paris
  Cedex 13, France}
\author{Alan D.~Sokal}
\affiliation{Department of Physics, New York University,
  4 Washington Place, New York, NY 10003, USA}
\affiliation{Department of Mathematics, University College London,
  London WC1E 6BT, UK}
\author{Andrea Sportiello}
\affiliation{Dip.~Fisica,
  Universit{\`a} degli Studi di Milano, and INFN, via G.~Celoria 16,
  20133 Milano, Italy}

\date{13 December 2008}
%%%%%%%%%%%%%%%%%%%%%%%%%%%% ABSTRACT
\begin{abstract}
\noindent
Consider the problem of determining whether there exists
a spanning hypertree in a given $k$-uniform
hypergraph. This problem is trivially in \mysf{P} for $k=2$, and
is 
\mysf{NP}-complete for $k\geq 4$, 
whereas for $k=3$, there exists a polynomial-time algorithm based on
Lov{\'a}sz' theory of polymatroid matching.

Here we 
give a completely different, randomized polynomial-time algorithm in the case $k=3$. 
 The main ingredients are a
Pfaffian formula by Vaintrob and one of the authors (G.M.)\ 
for a polynomial that enumerates spanning hypertrees with some signs, and a
lemma on the number of roots of polynomials over a finite field.
\end{abstract}

\maketitle

%%%%%%%%%%%%%%%%%%%%%%%%%%%%%%%%%%%%%%%%%%%%%%%%%%%%%%%
\section{Introduction}

\begin{figure}
\caption{\label{fig.hght}A 3-uniform hypergraph $G$, and a spanning
  hypertree $T \subset G$, whose edges are in darker gray.}
\includegraphics[scale=1, bb=2.5 0 247.5 160]{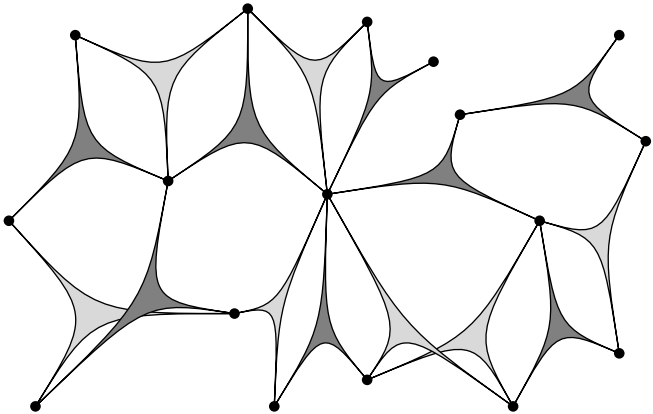}
\end{figure}

\noindent
A (finite) hypergraph $G=(V,E)$ consists of a finite set $V$ (the
\emph{vertex set}) and a set $E$ of subsets of $V$ (the
\emph{hyperedges}), each of cardinality at least 2.
We also write $V(G)=V$ and $E(G)=E$.
When all the
hyperedges have the same cardinality $k$, we say that the 
hypergraph is
\emph{$k$-uniform}.  In the case $k=2$ we are dealing with ordinary
(simple) graphs.
We say that $H \subseteq G$ is a {\em
sub-hypergraph} if $V(H) \subseteq V(G)$ and $E(H) \subseteq E(G)$; a
sub-hypergraph is {\em spanning} if $V(H)=V(G)$.  A hypergraph $H$ is
a {\em hypertree} if it is connected and there are no cyclic sequences
of vertices and hyperedges
\[
v_1, A_1, v_2, A_2, \cdots, v_{\ell}, A_{\ell}, v_{\ell+1}
\quad
( v_{\ell+1} = v_1 )
\]
such that $\ell \geq 2$, $v_i \neq v_{i+1}$ and $v_i, v_{i+1} \in A_i$.
An example of a 3-uniform hypergraph, together with 
a spanning sub-hypergraph that is a hypertree,
is shown in Figure~\ref{fig.hght}.

We will deal here with issues in Computational Complexity Theory
\cite{GJ}.  An introduction to the subject which includes the class
\mysf{RP} of probabilistic polynomial-time problems, pertinent to this
paper, can be found in Chapters 2--4 of Talbot and
Welsh~\cite{talbotwelsh}.

In this paper we are concerned with the following decision problem:

\begin{center}
\begin{minipage}[c]{7.5cm}
 {\sc $k$-Uniform Spanning Hypertree} (\sht{$k$}):
\\
 Given a $k$-uniform hypergraph $G=(V,E)$,
 determine whether there exists a spanning hypertree.
\end{minipage}
\end{center}

\noindent
Of course, every connected graph contains a spanning tree,
so \sht{$k$} is trivially in \mysf{P} for $k=2$
(it suffices to check whether $G$ is connected).
On the other hand, for $k \ge 3$ it is \emph{not} true that
every connected $k$-uniform hypergraph contains a spanning hypertree,
and the decision problem is highly nontrivial.

Our main result here is to provide an \mysf{RP} algorithm for
the $k$-Uniform Spanning Hypertree problem when $k=3$.
After a first version of this paper was completed, we have learnt from
Andras~Seb{\"o} that there is actually a polynomial-time algorithm for this
problem,
coming as a specialization of Lov{\'a}sz' algorithm for matching on
linear 2-polymatroids
\cite{Lov1,Lov2}. However, Lov{\'a}sz' techniques are completely different
from ours, and we believe that our more algebraic approach is of
independent interest. 
We remark that for $k\geq 4$ the spanning hypertree problem is
\mysf{NP}-complete by a result of C.~Thomassen which appears in \cite[Theorem 4]{af}. (We
thank Marc Noy for bringing this argument to our attention.)
Moreover, the same argument shows that the
corresponding {\em counting} problem is $\sharp$\mysf{P}-complete
already for $k \ge 3$.  
%%GM For the convenience of the reader, we 
We will
briefly review Thomassen's argument at the end of this
introduction. 

\vskip 8pt

{\em Organization of the paper.} The bulk of this paper has
two parts. In the first part, we discuss the
main ingredient of our \mysf{RP} algorithm, namely the Pfaffian-Tree Theorem
of \cite{masvain2} which expresses a signed version of the multivariate
spanning-tree generating function of a $3$-uniform hypergraph as a
Pfaffian.  Then in the second part, we describe our algorithm, first
intuitively, and then more formally,
and sketch the analysis of time- and space-complexity.
This part is fairly standard in complexity theory, but we hope that
the partly expository presentation of the various concepts involved
will be useful for the interdisciplinary audience (such as ourselves)
we have in mind. Finally, we end the paper with some speculations and
directions for further research suggested by our work.

\vskip 8pt

To conclude this introduction, here is, then, 
Thom\-assen's 
argument showing that \sht{k} is \mysf{NP}-complete for $k \geq 4$. 

We recall that an \emph{exact cover} in a hypergraph $G=(V,E)$
is a subset
$E' \subseteq E$ of the hyperedges such that every vertex of $G$ belongs to
exactly one hyperedge in $E'$. (In the special case where
$G$ is an ordinary graph, an exact cover is nothing but a perfect
matching.) 
Now consider the following decision problem:

\begin{center}
\begin{minipage}[c]{7.5cm}
 {\sc Exact cover by $k$-sets} (X$k$C):
\\
 Given a $k$-uniform hypergraph $G=(V,E)$,
 determine whether there exists an exact cover.
\end{minipage}
\end{center}

\noindent
X3C is known to be \mysf{NP}-complete,
and is classified as problem [\![SP2]\!] in
Garey and Johnson \cite{GJ}. (It is \mysf{NP}-complete even when
restricted to 3-partite hypergraphs, in this case being called
\emph{3-Dimensional Matching} (3DM, [\![SP1]\!]).)
Implicitly, analogous statements hold as well for any $k \geq 3$.
Conversely, X2C is polynomial, by matching techniques, even in its
optimization variant, e.g.~by Gallai-Edmonds algorithm
(see~\cite{LovPlu}). On the other hand, the corresponding counting problem (equivalently: counting perfect matchings on arbitrary
graphs),  is
known to be $\sharp$\mysf{P}-complete (Valiant \cite{V}).

Now, given 
an arbitrary
$k$-uniform hypergraph $G$, Thomassen constructs a
$(k+1)$-uniform 
hypergraph $G'$ as follows: add an extra vertex $\star$, and
let
\[
E(G')=\{e\cup \{\star\}\, |\, e\in E(G)\}
\ef.
\]
The key
observation is that spanning hypertrees of $G'$ correspond bijectively
to exact covers of $G$
(with the obvious bijection, namely deleting $\star$ from each hyperedge).
Thus, any algorithm for 
\sht{$(k+1)$}
provides an algorithm for X$k$C.
In other words, X$k$C is reducible to \sht{$(k+1)$}.

{}From what we said above about X$k$C, it follows that \sht{$k$} is
\mysf{NP}-complete for $k \geq 4$, and counting spanning hypertrees in
a 3-uniform hypergraph is $\sharp$\mysf{P}-complete,
as asserted. 
\vskip 8pt

%%%%%%%%%%%%%%%%%%%%%%%%%%%%%%%%%%%%%%%%%%%%%%%%%%%%%%%
\section{A Pfaffian formula}

\noindent
Let $G=(V,E)$ be a finite hypergraph on $N$ vertices. The multivariate
generating
function for spanning hypertrees on $G$ is
 \be
\label{eq.Znew}
Z_{G}(\vec{w}) =
\sum_{T \in \mathcal{T}(G)}
\prod_{A \in E(T)} w_A
\ef,
\ee
where $\mathcal{T}(G)$ is the set of spanning hypertrees of $G$, and 
the $ w_A$ are commuting indeterminates, one for each hyperedge $A\in
E(G)$. 

Assume that $G$ is $k$-uniform.  If $G$ has a spanning hypertree, then 
necessarily 
\be 
\label{eq.N}
N=(k-1)n+1~,
\ee  where $n$ is the number of hyperedges in each spanning hypertree.
 Therefore we will assume (\ref{eq.N}) from now on. 

It will be convenient to consider $G$ as a
sub-hypergraph of $\mathcal{K}(N,k)$, the complete $k$-uniform hypergraph
on $N$ vertices, which has an hyperedge for all unordered $k$-uples $A=\{ i_1, \ldots, i_k \} \subseteq [N]
:=\{1,2,\ldots ,N\}$. 
We denote
$Z_{\mathcal{K}(N,k)}(\vec{w})$
by $Z_{n,k}(\vec{w})$,
and $\mathcal{T}(\mathcal{K}(N,k))$ by $\mathcal{T}_{n,k}$.
Note that the degree of $Z_{n,k}(\vec{w})$ is
$n$, as given in
(\ref{eq.N}). The spanning hypertree generating function
$Z_{G}(\vec{w})$ of an arbitrary $k$-uniform hypergraph $G$ can be
obtained from  $Z_{n,k}(\vec{w})$ by setting the weights $w_A$
to zero for hyperedges $A$ not in $G$.

A classical result by Kirchhoff is that, for $k=2$, the expression
$Z_{n,2}(\vec{w})$ (and therefore $Z_G(\vec w)$ for any graph $G$)  is
given by a determinant. Defining the $N \times N$
\emph{Laplacian matrix} $L$ as
\be
L_{ij} = \left\{
\begin{array}{ll}
-w_{ij} & i \neq j \\
\sum_{k \neq i} w_{ik} & i = j
\end{array}
\right.
\ee
and taking a whatever $(N-1)$-dimensional principal minor $L(i_0)$
(i.e.~with row and column $i_0$ removed -- and remark that $N-1=n$ if
$k=2$), one has
\begin{theorem}[Matrix-Tree] % -- Kirchhoff
\be
\label{eq.MT}
Z_{n,2}(\vec{w}) = \det L(i_0)
\ef.
\ee
\end{theorem}

As is well-known, Kirchhoff's formula allows one to count spanning trees on a
graph $G$:  putting $w_A=1$ if $A \in E(G)$ and $w_A=0$ otherwise, the
determinant (\ref{eq.MT}) gives the cardinality of the set $\mathcal{T}(G)$
of spanning trees on~$G$. For later use, we remark that this formula also shows that counting spanning trees on
graphs is in \mysf{P}, as determinants can be evaluated in polynomial
time.  

For general $k$-uniform hypergraphs, no such formula is
known if $k\geq 3$. Moreover, a determinantal expression for
$Z_{n,k}(\vec{w})$ is unlikely to exist, since counting spanning
hypertrees is $\sharp$\mysf{P}-complete if $k\geq 3$, as discussed in the
introduction. 

We remark in passing that, nevertheless, the number of spanning
hypertrees on the complete hypergraph $\mathcal{K}(N,k)$ is known:
generalizing the classical result by Cayley for $k=2$, one has 
that
\be
\label{eq.Tnk}
|\mathcal{T}_{n,k}|
=
\frac{\big( (k-1)n \big)!}{\big( (k-1)n+1 \big) \, n!}
\left( \frac{(k-1)n+1}{(k-1)!} \right)^n
\ef,
\ee
as can be found in \cite{noibedini} (see also references therein).

{}From now on, we will consider the case $k=3$. Our
starting point is a 
recent result, due to A.~Vaintrob and one of
the authors (G.M.)
\cite{masvain2}, which states that an alternating sign version of the
spanning hypertree generating function $Z_{n,3}(\vec{w})$ is given by a Pfaffian.

We will denote this modified polynomial by
$Z^{\star}_{n,3}(\vec{w})$. For a given ordering of the vertices of the hypergraph (or,
equivalently, a labeling of the vertices with integers from $1$ to
$N$), a sign function $\epsilon(T)$ for hypertrees $T$ is defined
(more details are given later); then  
\be
\label{eq.preMV}
Z^*_{n,3} (\vec{w}) =
\sum_{T \in \mathcal{T}_{n,3}}
\epsilon(T)
\prod_{A \in E(T)} w_A
\ef.
\ee
To state the result, let $\epsilon_{ijk}$ be the totally antisymmetric tensor
(i.e. $\epsilon_{ijk}=0$ if two or more indices are equal,
$\epsilon_{ijk}=1$ if $i<j<k$ or any other cyclic permutation, and
$\epsilon_{ijk}=-1$ if $i<k<j$ or any other cyclic permutation) and define a
$N$-dimensional antisymmetric matrix $\Lambda$, with off-diagonal
elements
\be
\label{eq.defLambda}
\Lambda_{ij} =
\sum_{k\neq i,j} \epsilon_{ijk} w_{\{i,j,k\}}~.
\ee
Then one has, for any principal minor $\Lambda(i_0)$:
\begin{theorem}[Pfaffian-Hypertree~\cite{masvain2}]
\be
\label{eq.MV}
Z^*_{n,3} (\vec{w}) =
(-1)^{i_0 -1} \, \Pf \Lambda(i_0)
\ef.
\ee
\end{theorem}

\noindent
(The original formulation in \cite{masvain2} uses indeterminates
$y_{ijk}$ which are antisymmetric
in their indices. The correspondence with our notation here is simply that
$y_{ijk}=\epsilon_{ijk} w_{\{i,j,k\}}$.)

In order to give meaning to equations (\ref{eq.preMV}) and
(\ref{eq.MV}), we must
define the sign
function $\epsilon(T): \mathcal{T}_{n,3} \to \{\pm 1\}$.
Several equivalent definitions exist, and all of them
require making some arbitrary choices;
the proof that the resulting sign is actually independent of these
choices (developed to full extent
in~\cite{masvain2})
can be performed
inductively in tree size, studying the invariances in the elementary
step of adding a `leaf' edge to a tree.

It is worth stressing, however, that the precise determination 
of this
sign
function is \emph{not} 
actually used in  
our algorithm (and not
even in our proofs of complexity bounds).

The first definition of $\epsilon(T)$ given in \cite{masvain2} is
as follows.
For $S \subseteq [N]$, call
$\tau_S$ the permutation which rotates cyclically the elements of $S$
(in their natural order), and keeps fixed the others. Then, for a
given ordered $n$-uple of hyperedges $(A_1, \ldots, A_n)$ forming a tree
$T$, define the permutation $\hat{\tau} = \tau_{A_1} \cdots
\tau_{A_n}$. This permutation is composed of a single cycle of length
$N$. It is thus conjugated to the ``canonical'' $N$-cycle
$\tau_{[N]}$, i.e.~there exists $\sigma$ such that
\be
\label{eq.signconj}
\hat{\tau} = \sigma \, \tau_{[N]} \, \sigma^{-1}
\ef.
\ee
Then actually the signature $\epsilon(\sigma)$ does not depend on the
ordering of the hyperedges, but only on $T$, and
taking $\epsilon(T) := \epsilon(\sigma)$ is a valid definition,
and the appropriate one for (\ref{eq.MV}) to
hold.

An equivalent definition is as follows. Consider a planar embedding
of the tree, in such a way that for each hyperedge $A_\alpha$, if
$(i_{\alpha}, j_{\alpha}, k_{\alpha})$ denote its three vertices
cyclically ordered in the clockwise order given by the embedding, one
has $\epsilon_{ijk}=1$. (Such an embedding always exists.) Then
construct the string $\hat{\rho}$ of $3n$ symbols in $[N]$ (recall
that $n$ is the number of hyperedges) corresponding to the sequence of
vertices visited by a clockwise path surrounding the tree, starting
from an arbitrary vertex. Each of the $N$ vertices occurs in this
string, but some vertices appear more than once. Now remove
entries from this string until each of the $N$ vertices appears
exactly once, thus getting a string $\rho$ of $N$ distinct elements of
$[N]$. If we interpret this string as a permutation, then $\epsilon(T)
= \epsilon(\rho)$, despite of the arbitrariness of the choices for the
planar embedding, the starting point and the extra entries we choose
to remove. An example is given in \cite[Figure 3.1]{masvain2}.

Instead of using a planar embedding, one can also describe this
procedure by choosing a given vertex as root (say, $i_0$), and
orienting the edges accordingly (so that each edge has a ``tip'' and
two ``tail'' vertices, and all edges are oriented towards the
root). Now the vertices
inside any
hyperedge $A_{\alpha}$ can be
uniquely ordered $(i_{\alpha}, j_{\alpha}, k_{\alpha})$ such that
$i_{\alpha}$ is the tip and
$\epsilon_{i_{\alpha}j_{\alpha}k_{\alpha}}=1$. With this notation, the
sign $\epsilon(T)$ coincides with the sign of a certain monomial in
the $N$-dimensional exterior algebra
(or \emph{real Grassmann Algebra}):
\be
\label{eq.signexte}
e(i_0) \wedge
\Big(
\bigwedge_{\alpha}
e(j_{\alpha}) \wedge e(k_{\alpha})
\Big)
=
\epsilon(T)
\;
e(1) \wedge \cdots \wedge e(N)
\ef.
\ee
In order to see the equivalence with the previous definition of $\epsilon(T)$,
it suffices to take an appropriate planar embedding of $T$ and to
observe that we can perform the removal of extra entries
in such a way that in the final string $\rho$, the two tails of each
hyperedge always come consecutively. The ordering of the hyperedges in the
product on the L.H.S. of (\ref{eq.signexte})
is irrelevant at sight, as an hyperedge with odd size has an even
number of tails, which thus are commuting expressions in exterior
algebra. Nonetheless the true invariance is stronger, as it concerns also
the choice of the root vertex, and much more, as implied by the preceeding
paragraphs.
The expression (\ref{eq.signexte}) above is an efficient
way of computing $\epsilon(T)$ for a given tree $T$.

This last definition for $\epsilon(T)$ has also the advantage of
driving us easily towards a proof using Grassmann variables of formula
(\ref{eq.MV}). In fact, for both equations (\ref{eq.MT}) and
(\ref{eq.MV}), say with $i_0=1$, we can recognize the expressions for
Gaussian integrals of Grassmann variables, ``complex'' and ``real'' in
the two cases respectively
\begin{align}
Z_{n,2}(\vec{w})
&=
\int
\mathcal{D}(\psi, \psibar)
\; \psibar_1 \psi_1 \,
e^{
\sum%_{i<j}
w_{ij}
(\psibar_i - \psibar_j)
(\psi_i - \psi_j)
}
\ef;
\\
Z^*_{n,3}(\vec{w})
&=
\int
\mathcal{D}(\theta)
\; \theta_1 \,
e^{
\sum%_{i<j<k} \!
w_{ijk}
(\theta_i \theta_j + \theta_j \theta_k + \theta_k \theta_i)
}
\ef.
\end{align}
It is combinatorially clear that such an integral will generate an
expansion in terms of spanning subgraphs, with one component being a
rooted tree and all the others being unicyclic. Then one
realizes that, because of the anticommutation of the variables,
unicyclics get contributions with opposite signs, which ultimately
cancel out. However this mechanism occurs in slightly different ways
in the two cases (cfr.~\cite{usPRL} and \cite{abdes} respectively for
more details).

{\em Remark.} The original proof of the Pfaffian-hypertree theorem in
\cite{masvain2} was by induction on the number of hyperedges using a
contraction-deletion formula. Another proof was given in
\cite{masvain} which exploits the knot-theoretical context which
originally lead to the discovery of the formula. The proof using
Grassmann variables alluded to above is due to Abdesselam
\cite{abdes}. Finally, yet another proof
was given by
Hirschman and Reiner \cite{HR} using the concept of a sign-reversing
involution.

%%%%%%%%%%%%%%%%%%%%%%%%%%%%%%%%%%%%%%%%%%%%%%%%%%%%%%%

\section{Towards a Randomized Polynomial-time algorithm}

\noindent
Equation (\ref{eq.preMV}) provides us with a \emph{multivariate}
alternating-sign generating function, which is a polynomial, and
identically vanishing if and only if our hypergraph $G$ has no spanning
hypertrees.
Can we use this to decide efficiently whether $G$ has a spanning
hypertree?

Recall that in the classical Matrix-Tree theorem for
ordinary graphs there
are no signs, and upon putting edge weights equal to $1$ for edges in
the graph, and equal to zero otherwise, we get the number of spanning
trees on our graph. Moreover, calculating numerically the determinant
of a matrix can be
performed by Gaussian elimination in polynomial time.

Let us apply the same idea to the Pfaffian-Hypertree formula
(\ref{eq.MV}). Since $\det \Lambda(i_0)=(\Pf \Lambda(i_0))^2$, we
can again evaluate $Z^*_{n,3} (\vec{w})$ in polynomial time, if, as
before, we set
the hyperedge weights equal to $1$ for hyperedges in
a $3$-uniform hypergraph $G$, and equal to zero otherwise. But because of the signs
in $Z^*_{n,3}$,
this is not the number of spanning hypertrees in $G$!

One way out is to evaluate $\det \Lambda(i_0)$ at some {\em random} set of
numerical weights (but keeping weights equal to zero for hyperedges
not in the hypergraph). If one gets a non-zero result, this would
certainly prove that the multivariate generating function is non-zero,
and hence provide a certificate of the fact that a spanning hypertree
exists. Conversely, if many evaluations at random independent points
are zero, one starts believing that the graph has no hypertrees at
all. This na{\"\i}ve idea can be formalized within the framework of
the \mysf{RP} complexity class.

In complexity theory, the class of Randomized Poly\-no\-mi\-al-time problems
(\mysf{RP}) contains problems for which, given any instance, a
polynomial-time {\em probabilistic} algorithm can be called an
arbitrary number of times in such a way that:
\begin{itemize}
\item If the correct answer is `False', it always returns `False';
\item If the correct answer is `True', then it returns `False' for the
$t$-th query with a probability at most $1/2$, regardless of the
previous query results.
\end{itemize}

\noindent
We now discuss in more detail how to construct an \mysf{RP}-algorithm from
the Pfaffian-Hypertree formula
(\ref{eq.MV}).
Note that, of course, one has
\mysf{P}\,$\subseteq$\,\mysf{RP}\,$\subseteq$\,\mysf{NP}.

%%%%%%%%%%%%%%%%%%%%%%%%%%%%%%%%%%%%%%%%%%%%%%%%%%%%%%%
\section{Gaussian algorithm over finite fields}

Calculating numerically the determinant of a matrix using Gauss
elimination is commonly thought to be of polynomial time complexity
(at most cubic). If this is certainly true for ``float'' numbers (but
suffers from numerical approximations), some remark is in order for
``exact'' calculations. Indeed, in this case we have to choose a
field, such as $\mathbb{Q}$, which is suitable for exact numerical
computation through a sequence of sums, products and inverses, which,
in the complexity estimate above, have been considered an ``unity of
complexity'' (a variant is possible, in which one works in the ring
$\mathbb{Z}$ and recursively factors out a number of g.c.d.'s). This
is however not an innocent assumption. For example, if one works with
rational numbers with both numerators and denominators having a
bounded number of digits $d$, in general, during the Gauss procedure,
one may suffer from an exponential growth of this number (as the
l.c.m.~of two $d$-digit integers may well have $2d$ digits).

An improved choice, and which makes the analysis simpler, is to
consider finite fields $\GF_q$, for which the complexity of operations
is uniformly bounded. A primer in finite fields can be found, for
example, in the textbook~\cite{bookGF}. The field $\GF_q$ exists for
$q$ a power of a prime, $p^h$, as a quotient of a set of polynomials
with coefficients in $\mathbb{Z}_p$ by a polynomial of degree $h$,
irreducible in $\mathbb{Z}_p$. Also for finite fields, polynomials of
degree $n$ have at most $n$ distinct roots. Recall that $a^2 = 0$ iff
$a=0$ in any field, so that also in $\GF_q$ one has $\det \Lambda = 0$
only if
$\Pf \Lambda = 0$.

Two specially used cases are $q$ prime, in which $\GF_q$ just
coincides with $\mathbb{Z}_q$, equipped with the product modulo $q$,
and $q$ a power of 2, which is mostly used in Coding Theory. In both
cases, the arithmetic operations $+(\cdot,\cdot)$,
$\times(\cdot,\cdot)$ and $1/(\cdot)$ are performed in a very
efficient way, if a `Space' requirement of order $q$ is allowed for
storing a table of
discrete
logarithms (this will turn out to be subleading for
our problem). As a result,
if a given value for $q$ is chosen, calculating the determinant of a
$N$-dimensional matrix in $\GF(q)$ takes a time of order
$N^3 (\log q)^2$ (improved to $N^3 \log q$ if
a fast preprocessing, concerning the table of logarithms,
is performed
\cite{bookGF,bookShoup}). Remark that for us it is important to
control the dependence not just on $N$ but also on $q$, because 
in our context
$q$ will depend on~$N$.

%%%%%%%%%%%%%%%%%%%%%%%%%%%%%%%%%%%%%%%%%%%%%%%%%%%%%%%
\section{Roots of polynomials over a finite field}

\noindent
The analysis of the previous sections naturally induces us to study
the roots of polynomials over Galois fields~$\BF_q$.  In particular we
need an upper bound as follows:

\begin{lemma}
\label{lem.zeroes}
Let $f(x_1, \ldots, x_n)$ be a non-zero polynomial in $n$
variables of total degree $d$ with coefficients in the finite field
$\BF_q$. Then $f$ has at most $d q^{n-1}$ roots.
In other words, the probability that a randomly chosen $(a_1, \ldots,
a_n)\in \BF_q^n$ is a root of $f$ is $\leq d/q$.
\end{lemma}

\noindent
For the convenience of the reader, we include a proof of this lemma. An essentially
equivalent proof formulated in the language of
probability theory is given in 
\cite[Thm.~4.2]{talbotwelsh}.

Recall that 
the field
$\BF_q$ has $q$ elements. We say that
$(a_1, \ldots, a_n) \in \BF_q^{n}$ is a {\em non-root} of $f$ if
$f(a_1, \ldots, a_n)\neq 0$. We must show that $f$ has at least
$(q-d)q^{n-1} $ non-roots.

The proof is by induction on $n$, the number of variables.
For $n=1$,
a non-zero one-variable polynomial of degree $d$ has at most $d$
roots. Now assume $n\geq 2$.
Write
\[
f(x_1, \ldots, x_n)
=
\sum_{i=0}^m x_1^i \, g_i(x_2, \ldots, x_n)
\ef~,
\]
where $g_m\neq 0$
and $m \leq d$.
Observe that $g_m$ has degree at most $d-m$. By the induction
hypothesis applied to $g_m$, we know that $g_m$ has at most
$(d-m)q^{n-2}$ roots, and therefore we can find at least
$(q-d+m)q^{n-2}$ non-roots of $g_m$. For every $(a_2, \ldots, a_n)$
which is a non-root of $g_m$, consider
\[
f(x_1,a_2, \ldots, a_n)
\ef.
\]
This is a non-trivial polynomial in $x_1$ of degree exactly $m$,
therefore it has at most $m$ roots and at least $q-m$ non-roots. Thus
we have found at least $(q-d+m)q^{n-2}(q-m)$ non-roots of $f$ (namely
$ (q-d+m)q^{n-2}$ possibilities for $(a_2, \ldots, a_n)$ times $q-m$
possibilities for $a_1$.) It is easy to check that $(q-d+m)q^{n-2}
(q-m)\geq (q-d)q^{n-1}$. This completes the proof.
\hfill $\square$

%%%%%%%%%%%%%%%%%%%%%%%%%%%%%%%%%%%%%%%%%%%%%%%%%%%%%%%
\section{The algorithm}

\noindent
The lemma \ref{lem.zeroes} stated above implies that, for any
$3$-uniform hypergraph $G$ with $2n+1$ vertices, if one evaluates
$(Z_{n,3}^*)^2$ at some values $\{ w_{A} \}_{A \in E(G)}$ random
uniformly sampled from $\GF_q$ (and, of course, $w_A=0$ if $A \not\in
E(G)$), one gets always zero if $G$ has no spanning hypertrees, and
obtains zero although $G$ has some spanning hypertrees, with
probability at most $n/q$. If $q \geq 2n$, we are within the framework
of the \mysf{RP} complexity class.

So our algorithm is just as follows. Given $G$, choose a root vertex
and an ordering (once and for ever), and build the corresponding
matrix $\Lambda$ of indeterminates
as in (\ref{eq.defLambda})
(as a matrix of lists of edge-labels, with signs). Build the table of
$\GF_q$--logarithms for an appropriate value of $q$. Then, for a given
fault tolerance $\varepsilon=2^{-k}$, repeat $k$ times the
following probabilistic algorithm:
\begin{itemize}
\item extract the values $w_A$
independently identically distributed in
$\GF_q$;
\item evaluate $\Lambda$ in
numerical
form for these values;
\item evaluate the determinant numerically by Gauss elimination (in
$\GF_q$);
\item if the result is non-zero, return
``there are trees'', and break the cycle.
\end{itemize}
Then, if the cycle terminates without breaking, return
``there are no trees with probability $1-2^{-k}$\,''.

If we consider a generic value of $q > n$, when the cycle terminates
without breaking we know that there are no trees with probability
at least $1-(n/q)^{k}$.
For a given value of $\varepsilon$, we thus get an upper bound on the
time complexity (up to a multiplicative constant)
\[
n^3 \log q \,
\left\lceil
\frac{- \log \varepsilon}{\log q - \log n}
\right\rceil
\ef.
\]
If both $n, \varepsilon^{-1} \to \infty$, optimization in $q$ suggests
to take $q \sim n/\varepsilon$ and perform a single query,
(instead of taking
$q = 2n$ and performing
$\mathcal{O}(-\log \varepsilon)$ queries). This saves an
extra factor $\min(\log n, -\log \varepsilon)$, and gives a complexity
of order
$n^3 \max( \log n, -\log \varepsilon )$.

%%%%%%%%%%%%%%%%%%%%%%%%%%%%%%%%%%%%%%%%%%%%%%%%%%%%%%%
\section{Perspectives}

\noindent Given the datum of a 
$3$-uniform hypergraph $G=(V,E)$, a
positive integer $s$,
and a set of integer-valued edge costs $\{ c_A \}$, consider the cost
function for spanning hypertrees $\mathcal{H}(T) = \sum_{A \in E(T)}
c_A$. Then, one has the decision problem of determining if $G$ has any
spanning hypertree $T$ with $\mathcal{H}(T) \leq s$.
This problem is essentially equivalent to the corresponding
\emph{optimization} problem, of finding the spanning
hypertree of minimum cost,
and has a complexity at least as large as the one of \sht{$3$}.
At the time of this writing, we don't know whether Lov{\'a}sz'
polymatroid matching techniques can 
answer this problem, which is not addressed in \cite{Lov1,Lov2}. Be
that as it may, 
it would also be interesting to understand if an \mysf{RP}
algorithm exists for this problem,
through an extension of the
technique described in the present paper.

Another issue worth investigating may be how to use the exact
expression for the sign $\epsilon(T)$ and maximize the fraction of
terms getting the same sign in (\ref{eq.preMV}), in order to use the
Pfaffian-Hypertree formula in all of its strength. This may lead to a
notion of Pfaffian orientation for 3-uniform hypergraphs, analogous to
the notion of Pfaffian orientation (or Kasteleyn orientation) for
graphs \cite{RST}. Indeed, our story is somewhat parallel to the story
for perfect matchings
%AS-1212
%added
(and also Lov{\'a}sz algorithm is 
%%GM partially on the lines of 
a generalization of the
%%GM
Edmonds-Gallai algorithm for 
%%GM matching 
matchings 
%%GM
on graphs). 
In particular, if we construct a $3$-uniform
hypergraph $G'$ from a graph $G$ using the procedure described in the
introduction, the Pfaffian-Hypertree formula (\ref{eq.MV}) for $G'$
becomes the well-known alternating sign multivariate generating
function for perfect matchings of $G$ given by a Pfaffian. Now in the
graph case, there is the notion of Kasteleyn orientation which serves
to get rid of the signs in the Pfaffian and allows therefore to count
perfect matchings in this way. A Kasteleyn orientation exists for
planar graphs, but not in general. Is there an interesting class of
``Pfaffian orientable'' $3$-uniform hypergraphs for which the
Pfaffian-Hypertree formula can be used to count spanning hypertrees
exactly?

%%%%%%%%%%%%%%%%%%%%%%%%%%%%%%%%%%%%%%%%%%%%%%%%%%%%%%%
\section{Acknowledgements}

\noindent
We thank 
D.B.~Wilson 
and M.~Queyranne
for useful discussions, and M.~Noy
and A. Seb{\"o}
for valuable correspondence.
We also wish to thank the Isaac Newton Institute for Mathematical Sciences,
University of Cambridge, for generous support during the programme on
Combinatorics and Statistical Mechanics (January--June 2008),
where part of this work was carried out.

%%%%%%%%%%%%% BIBLIOGRAPHY %%%%%%%%%%%%%%%

\end{document}